\documentclass[aps,pra,twocolumn,amsmath,amssymb,nofootinbib,superscriptaddress]{revtex4}
\newcommand{\bra}[1]{\langle#1|}
\newcommand{\ket}[1]{|#1\rangle}

\usepackage[pdftex]{graphicx}
\usepackage{mathrsfs}
\usepackage[colorlinks]{hyperref}
\usepackage{braket}

\begin{document}

\bibliographystyle{plainnat}

%
%

\title{An introduction to boson-sampling}

%
%

\author{Bryan T. Gard}
\affiliation{Hearne Institute for Theoretical Physics and Department of Physics \& Astronomy, Louisiana State University, Baton Rouge, LA 70803, USA}

\author{Keith R. Motes}
\affiliation{Centre for Engineered Quantum Systems, Department of Physics and Astronomy, Macquarie University, Sydney NSW 2113, Australia}

\author{Jonathan P. Olson}
\affiliation{Hearne Institute for Theoretical Physics and Department of Physics \& Astronomy, Louisiana State University, Baton Rouge, LA 70803, USA}

\author{Peter P. Rohde}
\email[]{dr.rohde@gmail.com}
\homepage{http://www.peterrohde.org}
\affiliation{Centre for Engineered Quantum Systems, Department of Physics and Astronomy, Macquarie University, Sydney NSW 2113, Australia}

\author{Jonathan P. Dowling}
\affiliation{Hearne Institute for Theoretical Physics and Department of Physics \& Astronomy, Louisiana State University, Baton Rouge, LA 70803, USA}

\date{\today}

\frenchspacing

%
%

\begin{abstract}
Boson-sampling is a simplified model for quantum computing that may hold the key to implementing the first ever post-classical quantum computer. Boson-sampling is a non-universal quantum computer that is significantly more straightforward to build than any universal quantum computer proposed so far. We begin this chapter by motivating boson-sampling and discussing the history of linear optics quantum computing. We then summarize the boson-sampling formalism, discuss what a sampling problem is, explain why boson-sampling is easier than linear optics quantum computing, and discuss the Extended Church-Turing thesis. Next, sampling with other classes of quantum optical states is analyzed. Finally, we discuss the feasibility of building a boson-sampling device using existing technology. 
\end{abstract}

\maketitle

\section{Introduction}

\subsection{Motivation for linear optics quantum computing and boson-sampling}

To-date, many different physical implementations and models for quantum computing have been proposed. These implementations include atom and ion trap quantum computing, superconducting qubits, nuclear magnetic resonance, quantum dots, nuclear spin, and optical quantum computing. When describing an implementation, one can use various models of computation. These include the gate model 
\cite{nielsen}, cluster (or graph) states \cite{bib:Raussendorf01, bib:Raussendorf03}, topological, adiabatic \cite{farhi}, quantum random walks \cite{bib:ADZ}, quantum Turing machines \cite{bern}, permutational \cite{jordan2}, and the one-clean qubit models \cite{moussa}. The most familiar and intuitive model is the gate model as it is most analogous to the classical circuit model of computation. We use this gate model in order to describe linear optics quantum computing (LOQC) and eventually a special purpose subset, boson-sampling.

As stated, there exist many choices of implementations and computational models. But which model is likely to yield the first demonstrated quantum computer? The answer is likely not just one, but a composite of different choices for the various required components of a quantum computer. For this discussion, we will focus on LOQC for the allure of simple implementation of boson-sampling.

There is a long history in the physics community of investigations into the use of linear interferometers, particularly linear optics interferometers, as a type of quantum information processor. In most of the early research, the consensus was that a linear optics interferometer (alone) could not be used to make a universal quantum computer regardless of the input states. For example, in 1993 (a year before Shor's discovery of the now-famous quantum factoring algorithm) there appeared a paper by {\u C}ern{\' y} that proposed using a linear interferometer to solve \textbf{NP}-complete problems in polynomial time, but the scheme suffered from an exponential overhead in energy \cite{cerny}. Similarly, in 1996, Clauser \& Dowling showed that a linear optics Talbot interferometer could be used to factor integers in polynomial time but with either an exponential overhead in energy or physical size \cite{clauser}. Also in 1996, Cerf, Adami \& Kwiat showed how to construct a programmable linear optics interferometer that could perform any universal logic gate with single photon inputs.  This scheme too suffered an exponential overhead in spatial dimension. In 2002, Bartlett \emph{et al.} showed that even with quadratic nonlinearities any interferometer that processes only Gaussian state inputs can be efficiently simulated classically.  This comprised a continuous variable analog of the Gottesman-Knill theorem for discrete variables in the ordinary circuit quantum computation model \cite{bart}. 

This litany of no-go theorems led to the widespread belief that linear interferometry alone could not provide a path to universal quantum computation and that, as a corollary, all passive linear optics interferometers were thought to be efficiently simulatable on a classical computer. For completeness we will introduce the LOQC approach of Knill, Laflamme \& Milburn (KLM) \cite{knill,kok} in the following section, but the remaining focus of this chapter is instead on boson-sampling. This is because, for the KLM scheme's set of universal gates, one requires intermediate measurements on ancilla photons with a feed-forward mechanism that imparts a type of effective Kerr nonlinearity on the system \cite{lap}. We explicitly only discuss linear optics implementations due to the fact that present-day nonlinear Kerr media exhibit very poor efficiency \cite{nielsen} and very weak non-linearities.

It came as a surprise to many in the quantum optics community when Aaronson \& Arkhipov (AA) argued that, in general, the operation of a passive `linear' optics interferometer with Fock state inputs cannot likely be simulated by a classical computer \cite{aar}. In particular, if one samples the output distribution utilizing photon-number discriminating detectors, one cannot predict the outcome with a classical computer without an exponential overhead in time or resources. This has become known as the boson-sampling problem. 

Gard, \emph{et al.} independently reached the same conclusion in the context of trying to simulate multi-photon coincidence counts in the output of a linear optics implementation of a quantum random walk with multi-photon walkers \cite{gard2}. In follow up papers, Gard \emph{et al.} \cite{gard}, as well as Motes \emph{et al.} \cite{motes}, argued from a physical (as opposed to a computational complexity) point of view that this difficulty to simulate such interferometers arose from two necessary requirements: (1) The photons `interact' at the beamsplitters via a Hong-Ou-Mandel effect that gives rise to an exponentially large Hilbert space in the number-path degrees of freedom (ruling out a brute force simulation of the interferometer); and (2) That the simulation of the interferometer is tied to computing the permanent of a large matrix with complex entries, a problem known to be in the complexity class \textbf{\#P}-complete.  This complexity class is not only thought to be intractable for classical computers, but even for universal quantum computers \cite{bib:Ryser63}. While the first requirement is a necessary condition, it is not by itself sufficient to imply an intractable simulation. As a counterexample, the Gottesman-Knill theorem gives examples of quantum circuits where gates in the Clifford algebra class generate exponentially large amounts of qubit entanglement but are nevertheless classically simulatable. Since there are sometimes shortcuts through the exponential Hilbert space, by tying the simulation to the problem of permanent computation we expect it is very unlikely that any such shortcuts exist. In contrast, the equivalent sampling problem with fermions rather than bosons is known to be classically easy to simulate, as the problem relates to matrix determinants rather than permanents, which are known to be in the complexity class \textbf{P}, which is known to be efficiently classically simulatable.  \cite{gard}.

Since the first appearance of the AA paper in 2010 there has been an explosion of research into the field of boson-sampling. As we will discuss below, there have been a number of experiments utilizing three photons from spontaneous parametric down conversion (SPDC) sources \cite{ralph,broome,spring,anon,till,crespi} (although the validity of these experiments are under debate as not all three photons were heralded single photons \cite{bib:dowlingSchmampling}). The experimental work has continued in parallel to a number of theoretical developments considering the effects of loss, noise, decoherence, non-Fock inputs, scalability of SPDC sources, ion-trap implementations, and so forth \cite{rohde1,rohde2,jiang,motes,shch}. We will discuss and summarize these results and more in the sections below. 

Why is boson-sampling getting so much attention?  What is it good for? Boson-sampling is an example of a computationally complex mathematical problem that cannot be efficiently simulated on a classical computer, but with significantly reduced experimental requirements compared to universal quantum computing schemes. It is the first interesting example of a realistic post-classical computing paradigm, though the true scope and power of such machines is not yet fully understood. 

Is a passive linear interferometer good for anything other than simply implementing boson-sampling? The problem itself, other than being a computational curiosity, has no known practical applications or killer apps such as integer factorization. Prior to Shor's algorithm, the same question was asked of a universal quantum computer.  Feynman's work in the 1980s had hypothesized that an ordinary quantum computer could be used to carry out certain physics simulations without the exponential overhead required on a classical computer.  This hypothesis was not proved until Lloyd's work in 1996 \cite{feyn,lloyd}.

Whilst the first exponential speedup advantage for a quantum computer was the Deutsch-Jozsa algorithm, discovered in 1992, this problem also had no practical applications \cite{deutsch}. In many ways, the boson-sampling quantum computer is akin to the ordinary circuit-based quantum computer pre-Shor.  Perhaps passive linear optics interferometers, now that this hidden computational power has been uncovered, are good for something else besides the boson-sampling problem? This potential is what has captured the imagination of many researchers in the field. While in the following section we arrive at our boson-sampling scheme by way of LOQC, we still maintain several of its benefits.  Namely, there are no requirements for excessive cooling of the optical elements, we have long coherence times compared to the basic gate operations, and relatively simple to understand noise sources.

One final caveat -- in almost all papers on the topic of boson-sampling, the interferometer is described as a passive \emph{linear} device with non-interacting bosons (photons in this case). However, the Hong-Ou-Mandel effect (followed by a projective measurement) imparts an effective nonlinearity and hence an effective interaction at each beamsplitter. The presence or absence of a photon in one input mode radically changes the output state of a second photon in another input mode. This `interaction' between indistinguishable particles, known as the exchange interaction, arises simply from the demand that the multi-particle wavefunction be properly symmetrized. While not a `force' in the usual sense, it can give rise to quite noticeable effects. For example, the bound state of the neutral hydrogen molecule (the most common molecule in our Universe) arises from just such an exchange interaction. It is therefore a misnomer to describe these interferometers as linear devices with non-interacting bosons. The exchange interaction is just as real as tagging on an additional term in a Hamiltonian.  If one adds post-selection in the number basis to the mix, this imparts an effective Kerr-like nonlinearity between the bosons to boot \cite{lap}.   

\subsection{Introduction to linear optics quantum computing}
 
\begin{figure}[!htb]
 \includegraphics[width=0.5\columnwidth]{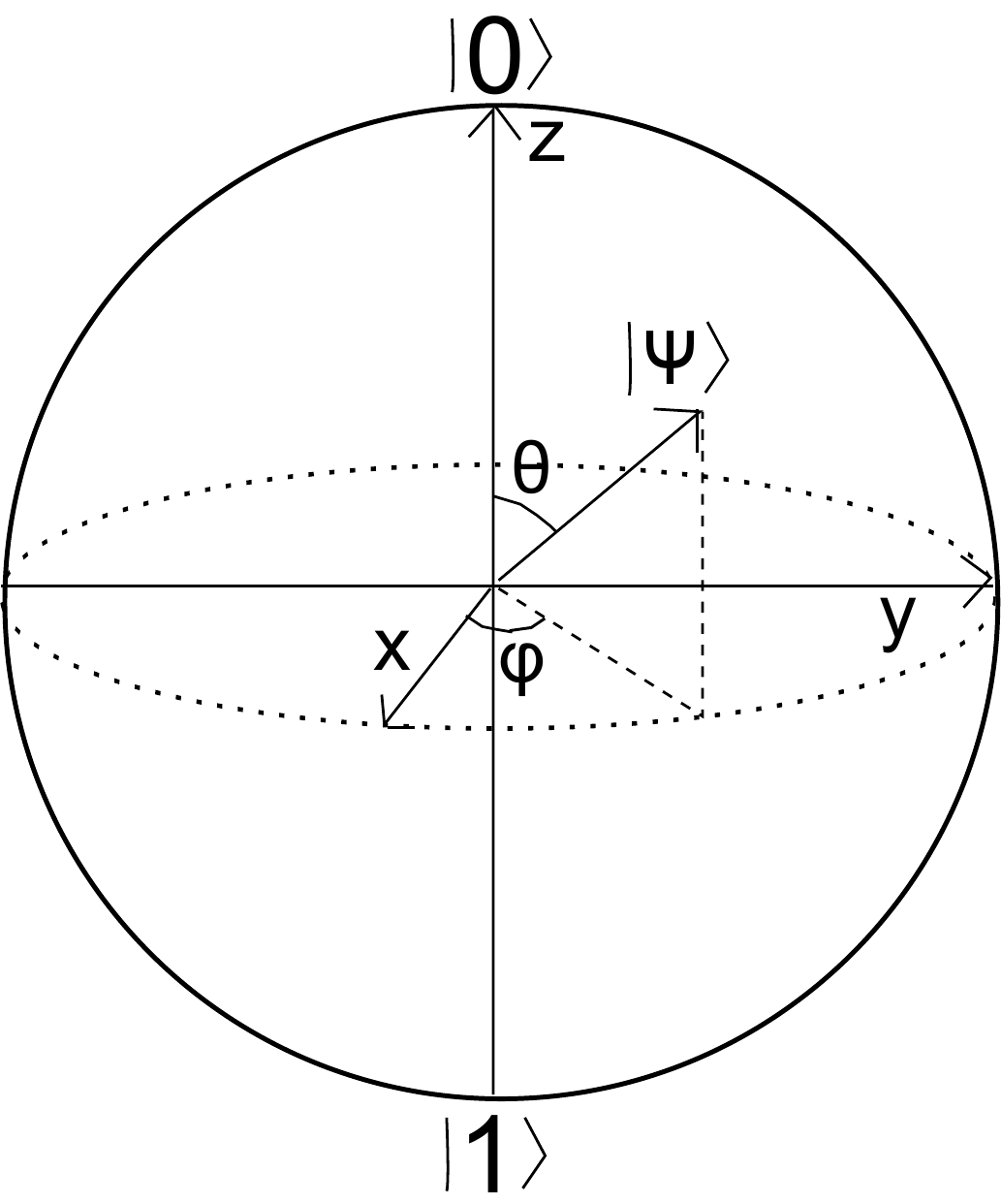}
 \caption{Bloch sphere showing a way to visualize the rotations that the Pauli matrices apply to a state. Pure states lie on the sphere while mixed states are contained within the sphere.}
 \label{fig:Bloch}
\end{figure}

 We begin by defining some terminology and notation. The smallest amount of data that we can deal with in quantum computing, analogous to the classical bit, is the quantum qubit. A qubit is defined as a unit vector in the complex two-dimensional vector space $\mathbb{C}^2$.  More simply, it can be represented in terms of the basis
\begin{eqnarray}
 \left[ \begin{array}{l}
1 \\
0 \\
\end{array} \right] = \ket{0},
\left[ \begin{array}{l}
0 \\
1 \\
\end{array} \right] = \ket{1}
\end{eqnarray}
for a zero and one qubit, respectively.  This constitutes a quantum superposition to which classical bits have no analog.  A general representation of a qubit is thus
\begin{equation}
\ket{\psi} = \cos\left(\frac{\theta}{2}\right) \ket{0} + e^{i \phi} \sin\left(\frac{\theta}{2}\right)\ket{1}.
\end{equation}
 If we let $\theta =\frac{\pi}{2}$ and $\phi =0$, we obtain a particular qubit of the form
\begin{equation}
\ket{\psi}= \frac{1}{\sqrt{2}}\left( \ket{0} +\ket{1} \right).
\end{equation}
In this superposition, our state (once measured) has a 1/2 probability of being in state zero and a 1/2 probability of being in state one. These superpositions can also be depicted on the Bloch sphere as shown in Fig.~\ref{fig:Bloch}. One may consider this superposition as being in both states at the same time.  However, once we measure the state in the logical basis, it collapses the superposition and takes either the value of zero or one. It is the act of measurement that forces the state to `choose' a zero or one. Thus from a classical perspective, there is an attribute of these superposition states to contain some `hidden' quantum information.

 Also analogous to classical computing, we need a set of logic gates to perform operations on our quantum states \cite{nielsen}. Some of the most common gates are defined as
\begin{eqnarray}
\textrm{Controlled-NOT (CNOT):} && \left[ 
\begin{array}{cccc}
1 & 0 & 0 & 0 \\
0 & 1 & 0 & 0 \\
0 & 0 & 0 & 1 \\
0 & 0 & 1 & 0 \\
\end{array} \right] \\ \nonumber
\textrm{Hadamard (H):} &&  \frac{1}{\sqrt{2}}\left[ 
\begin{array}{cc}
1&1 \\
1&-1 \\
\end{array} \right] \\ \nonumber
\textrm{Pauli-X ($\sigma_x$):} && \left[
\begin{array}{cc}
0&1\\
1&0\\
\end{array} \right] \\ \nonumber
\textrm{Pauli-Y ($\sigma_y$):} && \left[
\begin{array}{cc}
0&-i\\
i&0\\
\end{array} \right] \\ \nonumber
\textrm{Pauli-Z ($\sigma_z$):} && \left[
\begin{array}{cc}
1&0\\
0&-1\\
\end{array} \right] \\ \nonumber
\textrm{Phase:} && \left[
\begin{array}{cc}
1&0\\
0&i\\
\end{array} \right] \\
\frac{\pi}{8}: &\;& \left[
\begin{array}{cc}
1&0\\
0&e^{i \pi/4} \\
\end{array} \right]. \nonumber
\end{eqnarray}
The first of these, the CNOT gate, is a maximally entangling two-qubit gate, which is the quantum equivalent of the classical XOR gate.  The latter gates are single qubit gates, which implement rotations on the Bloch sphere. The single qubit gates may be trivially implemented using waveplates in quantum optics, whilst the CNOT gate is far more challenging, requiring an effective Kerr non-linearity.

These gates form one possible choice of a universal gate set. Any choice for a universal gate set can approximate any other gate set to arbitrary precision. So far, there are three classes of problems to which quantum computation outperforms classical computing. The first such class contains algorithms that make use of the quantum Fourier transform (such as Shor's algorithm for factoring and discrete logarithms). For $ N=2^n$ numbers, a classical fast Fourier transform would require $ N \log{N} \approx 2^n n$ steps while a quantum computer could do this same transform in only $\log^2{N} \approx n^2$ steps \cite{nielsen}.

Another class consists of quantum search algorithms which make use of superposition to speed search times. The most well known example of such an algorithm was discovered by Grover \cite{bib:grover}, where in a search of an unstructured database of $N$ elements, one wants to find an element of that search space satisfying a specific property. On a classical computer this search would require $O(N)$ operations, whilst a quantum search could accomplish this in $O(\sqrt{N})$ operations.

The third class is quantum simulation, where one simply attempts to simulate the evolution of a quantum system.  It is not a surprise that this class would require a quantum computer to simulate efficiently. For a classical computer to simulate a quantum system with $n$ distinct components, it would require $O(exp(n))$. A quantum computer would only require $O(n)$ qubits of memory however, where the proportionality constant depends on the choice of the physical system being simulated. We thus reduce an exponential resource use to only a linear one! For further discussion on quantum optics and quantum information processes see Refs. \cite{walls,aharonov,diVincenzo}.

In computational terms, a computation is considered efficient if the required resources and time scale at most polynomially with the size of the input. With only linear optical elements such as beamsplitters, phase-shifters, photodetectors, and feedback from photodetector outputs, it can be shown that one can achieve this efficiency. Using only linear optical elements, it can be shown that we can implement \cite{knill},
\begin{enumerate}
\item{Non-deterministic quantum computation.} 
\item{Probability of success of quantum gates approach unity.}
\item{Coding methods that achieve fault tolerance.} 
\end{enumerate}

Discussion of linear optics quantum gate efficiency, such as beamsplitters and the controlled phase gate are discussed in Ref.~\cite{lemr} with the description of entanglement power and entanglement efficiency.

\subsection{Linear optics quantum computing}

In general, to fully achieve a true quantum computer we require a way to prepare quantum states, perform a universal gate set on the qubits, and measure the output state.

In order to generate a quantum state we use a single photon source which adds a photon to the vacuum state $\ket{0}$ and thus sets any vacuum mode to the $\ket{1}$ state. This process is non-deterministic but is sufficient for quantum computing.

The simplest optical elements are phase-shifters and beamsplitters. These elements are used to act as gate operations on our prepared states. Since both of these transformations are unitary we can write each of these elements in terms of their unitary matrix. A phase-shifters unitary, acting on a single mode, is simply, $\hat{P}_{\phi}=e^{i \hat{n}\phi}$, where $\hat{n}$ is the number operator, while the unitary matrix for a beamsplitter is given by
\begin{eqnarray}
B_{\theta,\phi}=
\left(
\begin{array}{cc}
\cos{\theta}& -e^{i \phi} \sin{\theta} \\
e^{-i \phi} \sin{\theta}& \cos{\theta} 
\end{array}
\right),
\end{eqnarray}
in the basis of optical modes, where $\phi$ give phase relationships and $\theta$ stipulate the bias of the beamsplitter.


In order to measure the state, we use photodetectors which destructively determine if a mode contains a photon or not. For states with more than one photon then, we need a photon counting detector, which can be implemented by using a series of beamsplitters and photodetectors. The beamsplitters act so that the photons are spread evenly over $N$ modes, with each mode containing a photodetector. The probability of under-counting given that the photon number is $k$ is at most $k(k-1)/(2N)$. This is referred to as multiplexed photodetection \cite{bib:RohdeWebb07,bib:Fitch03,bib:Achilles04,bib:LPOR201400027,bib:ma2011experimental}. Another alternative is to use photon-number-resolving detectors.

In addition to these single qubit rotations we also require a nonlinear sign-flip (NS) gate \cite{knill}. This gate implements the transformation 
\begin{equation}
\mathrm{NS}: \alpha_0 \ket{0} + \alpha_1 \ket{1} +\alpha_2 \ket{2} \rightarrow \alpha_0 \ket{0} + \alpha_1 \ket{1} - \alpha_2 \ket{2},
\end{equation}
and is the basis of implementing the CNOT gate. This two qubit gate along with the previously discussed single qubit gates form the required universal gate set to perform quantum computing. One only needs a set of one- and two-qubit universal gates in order to construct general multi-qubit gates. Specifically we only require the Hadamard, phase, $\pi/8$ and CNOT gates \cite{nielsen}.

Using just linear optics and photodetection, implementing the NS gate is non-deterministic, which implies that with multiple gates in our circuit, the success probability of the computation drops exponentially with the number of gates. To overcome this, another useful tool in LOQC is the use of quantum gate teleportation to increase the probability of success of non-deterministic gates \cite{knill,bib:GottesmanChuang99}. Here we use two Bell pairs (maximally entangled two-qubit states) as a resource to teleport the action of a gate onto two qubits. This teleportation trick increases the success probability of the non-deterministic gate, but is itself non-deterministic. However, by concatenating the teleportation protocol we are able to increase the success probability of a non-deterministic gate asymptotically close to unity \cite{knill}, enabling efficient large-scale computation.

\subsection{Why is linear optics quantum computing hard?}

All of this may lead one to ask, if this scheme, using only linear elements is so simple, what's the hold up in implementing it? To implement this scheme we require a myriad of technicalities. These include synchronization of pulses, mode-matching, quickly controllable delay lines, tunable beamsplitters and phase-shifters, single-photon sources, and accurate, fast, single photon detectors. Most of this list is not terribly unrealistic to adhere to but current efficiencies of photodetectors are not at the point at which they may realistically implement the teleportation and the more complex gate operations (two qubit gates). The feedback control of these detectors must also be extremely fast in order to select proper state preparation before photon loss becomes an issue.

As an example, if we investigate actual implementation of a teleported (i.e high success probability) CNOT gate, which requires many individual non-deterministic CNOT gates, we can attain a probability of success in implementing this entangling operation of $95\%$ with approximately 300 successful CNOT gates which translates to an excessively large number ($>10^4$) of optical elements \cite{nielsen2}. Whilst this may seem daunting, recent approaches using cluster states have reduced experimental requirements by orders of magnitude \cite{bib:Nielsen04, bib:BrowneRudolph05}, but nonetheless the experimental requirements are substantial and still require challenging technologies such as fast-feedforward and dynamic control.

Without accurate implementation of these protocols we likely lose our claim to universality, but we still retain our ability to investigate some interesting problems. This realm of LOQC without fast feedback control or unrealistically accurate photodetectors lead us into boson-sampling.

\section{The boson-sampling formalism}

Unlike full LOQC, which requires active elements, the boson-sampling model is strictly passive, requiring only single-photon sources, passive linear optics (i.e beamsplitters and phase-shifters), and photodetection. No quantum memory or feedforward is required.

\begin{figure}[!htb]
\includegraphics[width=0.7\columnwidth]{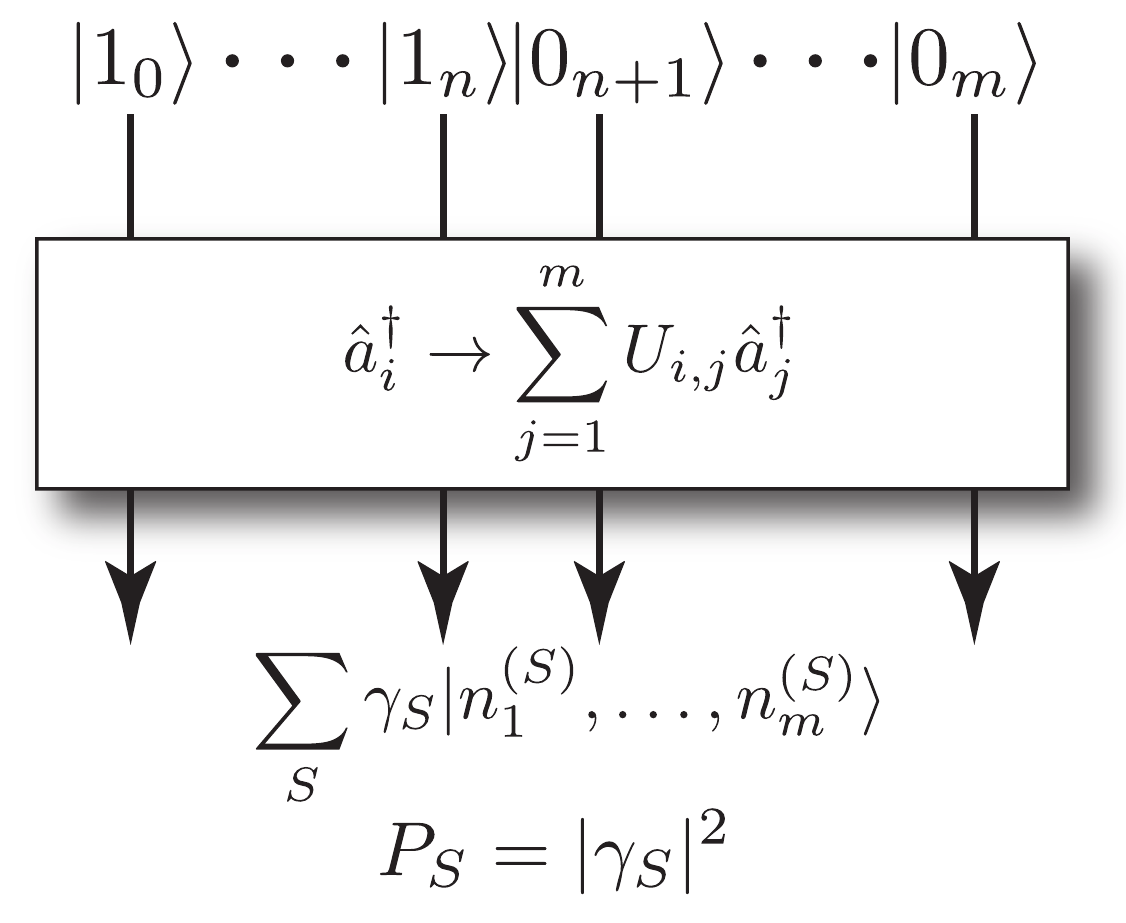}
\caption{The boson-sampling model. $n$ single photons are prepared in $m$ optical modes. These are evolved via a passive linear optics network $\hat{U}$. Finally the output statistics are sampled via coincidence photodetection. The experiment is repeated many times, reconstructing the output distribution $P_S$.} \label{fig:model}
\end{figure}

We begin by preparing an input state comprising $n$ single photons in $m$ modes,
\begin{eqnarray} \label{eq:input_state}
\ket{\psi_\mathrm{in}} &=& \ket{1_1,\dots,1_n,0_{n+1},\dots,0_m} \nonumber \\
&=& \hat{a}^\dag_1 \dots \hat{a}^\dag_n \ket{0_1,\dots,0_m},
\end{eqnarray}
where $\hat{a}^\dag_i$ is the photon creation operator in the $i$th mode. It is assumed that the number of modes scales quadratically with the number of photons, \mbox{$m=O(n^2)$}. The input state is evolved via a passive linear optics network, which implements a unitary map on the creation operators,
\begin{equation} \label{eq:Utransform}
\hat{U}\hat{a}_i^\dag\hat{U}^\dag = \sum_{j=1}^m U_{i,j} \hat{a}_j^\dag,
\end{equation} 
where $\hat{U}$ is a unitary matrix characterizing the linear optics network. It was shown by Reck \emph{et al.} \cite{bib:Reck94} that any $\hat{U}$ may be efficiently decomposed into $O(m^2)$ optical elements. The output state is a superposition of the different configurations of how the $n$ photons could have arrived in the output modes,
\begin{equation} \label{eq:outputState}
\ket{\psi_\mathrm{out}} = \sum_S \gamma_S \ket{n_1^{(S)},\dots,n_m^{(S)}},
\end{equation}
where $S$ is a configuration, $n_i^{(S)}$ is the number of photons in the $i$th mode associated with configuration $S$, and $\gamma_S$ is the amplitude associated with configuration $S$. The probability of measuring configuration $S$ is given by \mbox{$P_S = |\gamma_S|^2$}. The full model is illustrated in Fig.~\ref{fig:model}

It was shown by Scheel \cite{bib:Scheel04perm} that the amplitudes $\gamma_S$ are related to matrix permanents,
\begin{equation}
\gamma_S = \frac{\mathrm{Per}(U_S)}{\sqrt{n_1^{(S)}!\dots n_m^{(S)}!}},
\end{equation}
where $U_S$ is an \mbox{$n\times n$} sub-matrix of $U$, and \mbox{$\mathrm{Per}(U_S)$} is the permanent of $U_S$.

\begin{figure}[!htb]
\includegraphics[width=0.7\columnwidth]{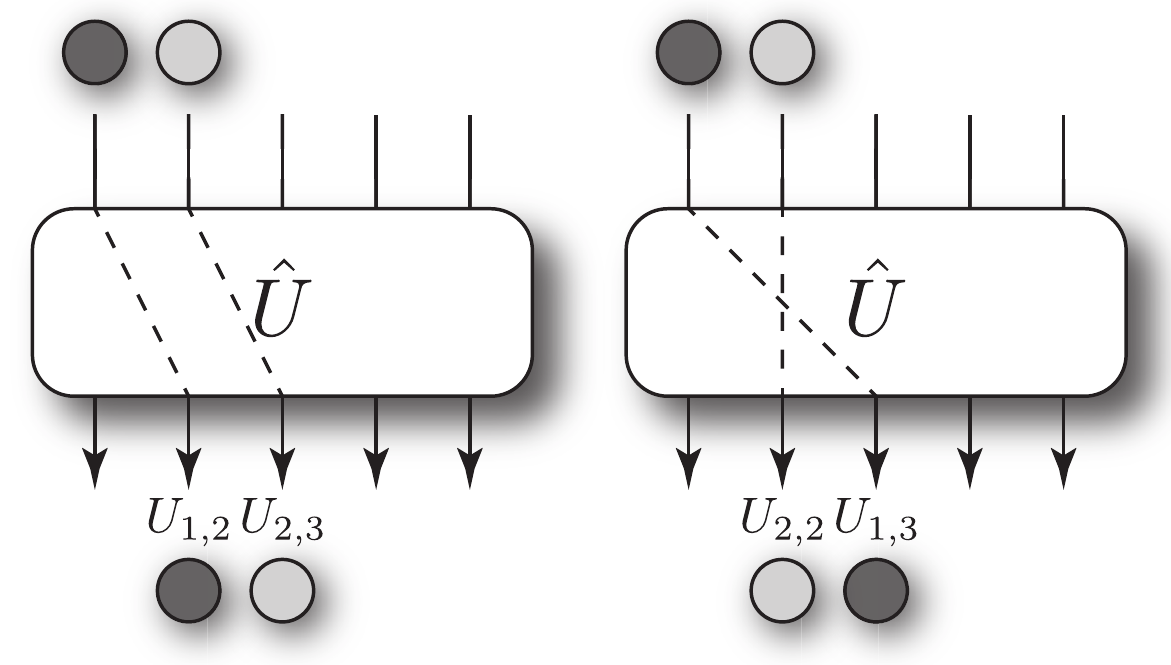}
\caption{Two-photon boson-sampling, where we wish to calculate the amplitude of measuring a photon at each of the output modes 2 and 3. There are two ways in which this may occur -- either the photons pass straight through, or swap, yielding a sum of two paths.} \label{fig:two_photon_perm}
\end{figure}

Let us examine this relationship with the permanent more closely. Consider Fig.~\ref{fig:two_photon_perm}. Here the first two modes have single photons, with the remaining modes in the vacuum state. Let us consider the amplitude of measuring one photon at output mode 2 and another at output mode 3. Then there are two ways in which this could occur. Either the first photon reaches mode 2 and the second mode 3, or vice versa, i.e the photons pass straight through, or swap. Therefore there are \mbox{$2!=2$} ways in which the photons could reach the outputs. Thus, this amplitude may be written as,
\begin{eqnarray} \label{eq:coinProbEx}
\gamma_{\{2,3\}} &=& \underbrace{U_{1,2}U_{2,3}}_{\mathrm{walkers\ don't\ swap}} + \underbrace{U_{1,3}U_{2,2}}_{\mathrm{walkers\ swap}} \nonumber \\
&=& \mathrm{Per} \left[ {\begin{array}{cc}
   U_{1,2} & U_{2,2} \\
   U_{1,3} & U_{2,3} \\
  \end{array} } \right],
\end{eqnarray}
which is a \mbox{$2\times 2$} matrix permanent.

\begin{figure}[!htb]
\includegraphics[width=\columnwidth]{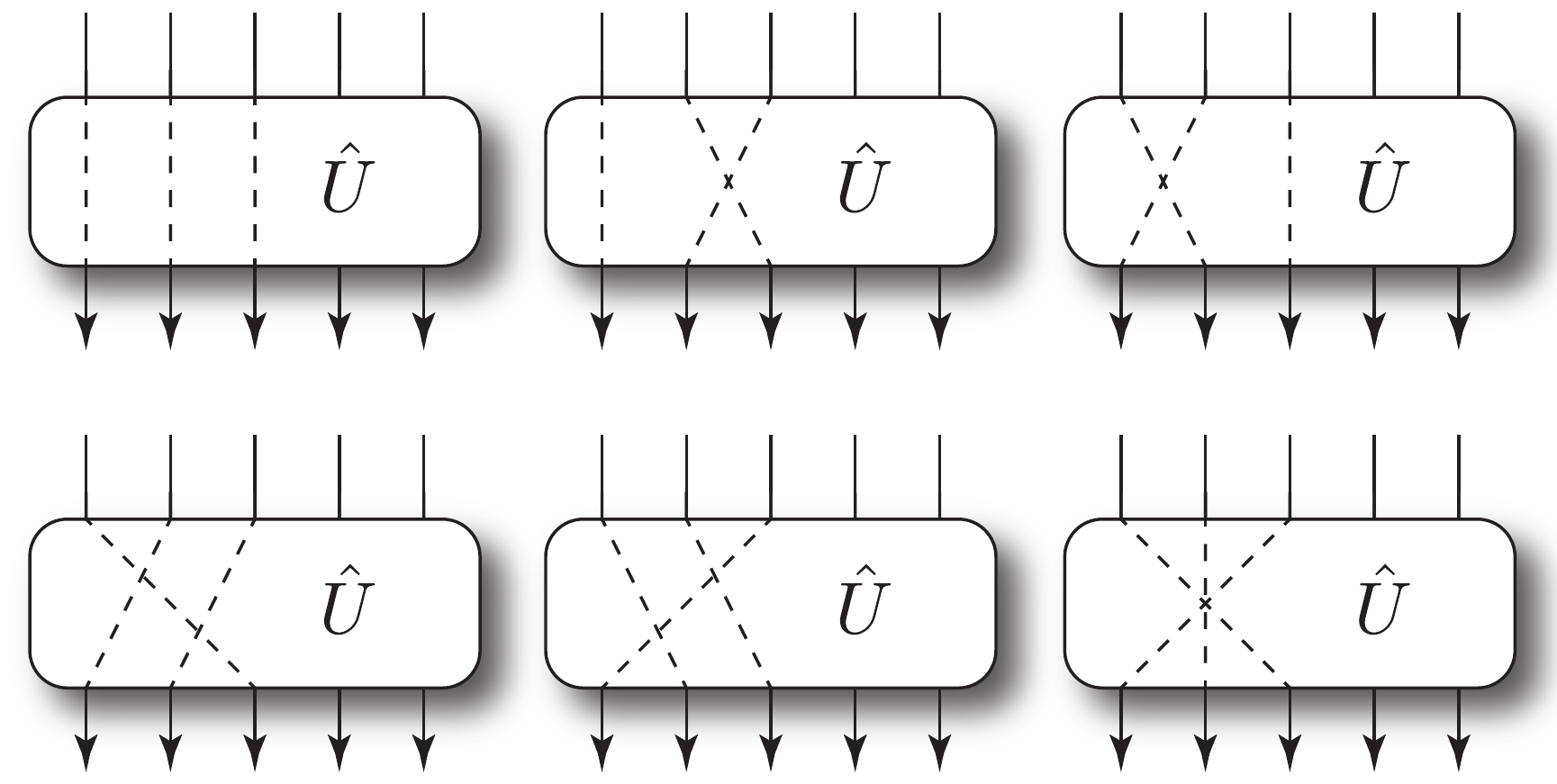}
\caption{Three photon boson-sampling, where we wish to calculate the amplitude of measuring a photon at each of the output modes 1, 2 and 3. There are now \mbox{$3!=6$} possible routes for this to occur.} \label{fig:three_photon_perm}
\end{figure}

As a slightly more complex example, consider the three photon case shown in Fig.~\ref{fig:three_photon_perm}. Now we see that there are \mbox{$3!=6$} ways in which the three photons could reach the outputs, and the associated amplitude is given by a \mbox{$3\times 3$} matrix permanent,
\begin{eqnarray} \label{eq:coinProbEx3}
\gamma_{\{1,2,3\}} &=& U_{1,1}U_{2,2}U_{3,3} + U_{1,1}U_{3,2}U_{2,3} \nonumber \\
&+& U_{2,1}U_{1,2}U_{3,3} + U_{2,1}U_{3,2}U_{1,3} \nonumber \\
&+& U_{3,1}U_{1,2}U_{2,3} + U_{3,1}U_{2,2}U_{1,3}
\nonumber \\
&=& \mathrm{Per} \left[ {\begin{array}{ccc}
   U_{1,1} & U_{2,1} & U_{3,1} \\
   U_{1,2} & U_{2,2} & U_{3,2} \\
   U_{1,3} & U_{2,3} & U_{3,3} \\
  \end{array} } \right].
\end{eqnarray}

In general, with $n$ photons, there will be $n!$ ways in which the photons could reach the outputs (assuming they all arrive at distinct outputs), and the associated amplitude will relate to an \mbox{$n\times n$} matrix permanent. Calculating matrix permanents is known to be \textbf{\#P}-complete, even harder than \textbf{NP}-complete, and the best known algorithm is by Ryser \cite{bib:Ryser63}, requiring \mbox{$O(2^n n^2)$} runtime. Thus, we can immediately see that if boson-sampling were to be classically simulated by calculating the matrix permanents, it would require exponential classical resources.

Because the number of modes scales quadratically with the number of photons, for large systems we are statistically guaranteed that all photons will arrive at different output modes. This implies that in this regime on/off (or `bucket') detectors will suffice, and photon-number resolution is not necessary, a further experimental simplification compared to full-fledged LOQC.

The number of configurations in the output modes scales as,
\begin{equation}
|S| = \binom{n+m-1}{n},
\end{equation}
which is super exponential is $n$. Thus, with an `efficient' (i.e polynomial) number of trials, we are unlikely to sample from a given configuration more than once. This implies that we are unable to determine any given $P_S$ with more than binary accuracy. Thus, boson-sampling does \emph{not} let us \emph{calculate} matrix permanents, as doing so would require determining amplitudes with a high level of precision, which would require an exponential number of measurements.

The experiment is repeated many times, each time performing a coincidence photodetection at the output modes. Thus, after each run we sample from the distribution $P_S$. This yields a so-called \emph{sampling problem}, whereby the goal is to sample a statistical distribution using a finite number of measurements. This is in contrast to well-known \emph{decision problems}, such as Shor's algorithm \cite{bib:Shor97}, which provide a well-defined answer to a well-posed question. Because boson-sampling is a sampling problem, finding a computational application is further complicated \textemdash if every time we run the device we obtain a different outcome, how does the outcome answer a well-defined question, and how do we map it to a problem of interest? This is one of the central challenges of boson-sampling \textemdash what can we do with it?

This sampling problem was shown by AA to be a computationally hard problem. That is, reconstructing the statistical distribution at the output to the boson-sampling device is computationally hard. However, whilst shown to be computationally hard, no known applications for boson-sampling have been described. Thus, boson-sampling acts as an interesting proof-of-principle demonstration that passive linear optics can outperform classical computers, but, based on present understanding, does not solve a problem of practical interest.

\subsection{Errors in boson-sampling}

There are a number of subtleties inherent in boson-sampling, often arising because one considers \textit{estimation} of a distribution.  In particular, we have shown that the amplitude of a given configuration of photons is tied to computing the permanent of a matrix.  While finding the exact permanent of a binary matrix is known to be \textbf{\#P}-complete, one can efficiently \textit{estimate} the permanent of a matrix with real, non-negative entries \cite{bib:Valiant79,bib:Jerrum04}.  If efficient estimation were also possible for complex-valued unitaries, one could pass off the boson-sampling problem as simply a type of singularity \textemdash a mathematical anomaly occurring only when one tries to know exact values. Since implementing a physical system such as boson-sampling is bound to have some kind of error, one could not hope to experimentally achieve the true distribution anyway.

In their paper, AA went to great lengths to prove the robustness of their result in the presence of error. Thus, even attempting to estimate the output distribution of a boson-sampling machine is likely computationally hard. (As one might expect, this correlates with the result that complex-valued matrix permanents cannot in general even be efficiently estimated unless exact permanents can as well \cite{bib:Jerrum04}.) In a sampling regime, one cannot tolerate so much error without eventually deviating too far from the desired distribution or requiring so many samples as to make the algorithm inefficient. For example, one could try postselecting to remove errant samples where photon-loss occurred.  However, if the probability were to scale on the order of $O(e^{-n})$, we would lose any hope of scaling for large $n$.

What then is an acceptable level of error? Suppose we fix an error threshold $\epsilon$ (i.e. we set $\epsilon$ to be the maximum allowable variation distance from the true distribution). Then so long as we have a success probably $P$ where $P>$ 1/poly($n$), we could correctly (within $\epsilon$) and efficiently sample a boson-sampling device with arbitrarily large photon number. More generally, if we wish also to scale $\epsilon$ smaller, $P>$ 1/poly($n,1/\epsilon$) \cite{aar}. As we discuss in the next section, however, we cannot experimentally determine whether this is achievable for asymptotic $n$.

\section{Boson-sampling and the Extended Church-Turing thesis}

Any model for quantum computation is subject to errors of some form. In the conventional circuit model, this includes errors such as dephasing. In linear optics, this includes photon loss and mode-mismatch. Let us consider a very generic error model for boson-sampling, where the single-photon states are the desired single photon with probability $p$, otherwise are in some erroneous state \cite{bib:BSECT}. This erroneous state could, for example, comprise terms with the wrong photon number (such as loss or second order excitations), or mode-mismatch. Then our input state is of the form,
\begin{equation} \label{eq:error_model}
\hat\rho_\mathrm{in} =\left(\bigotimes_{i=1}^n[p\ket{1}\bra{1} + (1-p)\hat\rho_\mathrm{error}^{(i)}]\right) \otimes [\ket{0}\bra{0}]^{\otimes^{m-n}},
\end{equation}
where \mbox{$\hat\rho_\mathrm{error}^{(i)}$} may be different for each input mode $i$. This is an independent error model, whereby each state is independently subject to an error channel. $p$ stipulates the fidelity of the single photon states. When \mbox{$p=1$}, the states are perfect single photons, and when \mbox{$p<1$}, the state contains erroneous terms. We desire to sample from the distribution of Eq.~\ref{eq:input_state}, whereby none of the input states are erroneous. This occurs with probability $p^n$.

Let $P$ be the probability that upon performing boson-sampling we have sampled from the correct distribution, otherwise we sample from noise. The complexity proof provided by AA only considered the regime where \mbox{$P>1/\mathrm{poly}(n)$}. Thus, for computational hardness, we require \mbox{$p^n > 1/\mathrm{poly}(n)$}. Clearly in the asymptotic limit of large $n$, this bound can never be satisfied for any $p<1$. Thus, with this independent error model, boson-sampling will always fail in the asymptotic limit.

Numerous authors \cite{bib:Broome20122012, bib:ShenDuan13, bib:AA13response, bib:Shchesnovich13, bib:Molmer13} have claimed that large-scale demonstrations of boson-sampling could provide elucidation on the validity of the Extended Church-Turing (ECT) thesis \textemdash the statement that any physical system may be efficiently simulated on a Turing machine. However, it must be noted that the ECT thesis is by definition an asymptotic statement about arbitrarily large systems. Because the required error bound for boson-sampling is never satisfied in this limit, it is clear that boson-sampling cannot elucidate the validity of the ECT thesis as asymptotically large boson-sampling devices must fail under an independent error model.

This concern might be overcome in the future with either (1) a loosening of the error bound to $1/\mathrm{exp}(n)$, or (2) the development of fault-tolerance techniques for boson-sampling. However, to-date no such developments have been made. Thus, based on \emph{present} understanding, boson-sampling will not answer the question as to whether the ECT thesis is correct or not. However, this is distinct from the question `will boson-sampling yield \emph{post-classical} computation?'. The answer to this question may very well be affirmative, as this only requires a finite sized device, just big enough to beat the best classical computers.

\section{Boson-sampling with other classes of quantum optical states}

Recall that the original proposal for boson-sampling by AA proceeds in three steps: 
\begin{enumerate}
\item{ Input - Vacuum and single photon Fock states.}
\item{ Evolve - Via a passive linear interferometer.}
\item{ Measure - Using an on-off photodetector.}
\end{enumerate}

Together, this process proves to be classically hard to simulate.  While this process is lauded for its relative simplicity compared to universal quantum computation, it is natural to ask if there are other systems that share these same properties.  That is, one can consider a generalization of boson-sampling by replacing one or more of the above three procedures with an analogous one.  The difficulty, as was the case with the original proposal, is proving that the system is classically intractable.

One could attempt to construct a complexity proof directly, but a rigorous proof may be hard to achieve; even AA's result for the approximate boson-sampling case relies on several likely yet unproven lemmas.  A simpler approach is to show that a system is equivalent to boson-sampling, implementing the same logical problem.  In other words, with only small overhead, the resulting statistics of one system could be used the compute the statistics of the other.  Informally, we might say a problem is `boson-sampling-hard' if it is at least as hard to compute as a boson-sampling distribution.  By showing this property of a system, one can extend AA's result to a more general class of problems.

One straightforward instance of such a proof was given by Seshadreesan \emph{et. al} \cite{bib:Pacs13} by generalizing the input to coherent states (instead of vacuum) and photon-added coherent states (instead of single photon Fock states).  More precisely, the input takes the form
\begin{equation}
\ket{\psi_{in}}=\hat{a}^\dag_1 \dots \hat{a}^\dag_n \ket{\alpha_1\dots\alpha_m}.
\end{equation}
Without changing the measurement scheme, intuitively one might expect this would work only when the coherent state has a small amplitude $|\alpha|$ since large amplitude coherent states tend to act `more classically' than other quantum states.  Additionally, it would likely become difficult to distinguish between photons contributed from a large amplitude coherent state and a single photon contributed from the creation operators.  Indeed, for coherent states the amplitude must scale according to $|\alpha|<1/\mathrm{poly}(n)$ for AA's result to hold.  A similar result holds for any general separable state close to vacuum.  It remains an open question whether other measurement schemes (such as homodyne detection) could be used to perform analogous boson-sampling on coherent or other quantum states.

Another class of states considered by Rohde \emph{et al.} \cite{bib:RohdeCat} was cat states -- superpositions of coherent states ($\ket{\alpha}+\ket{-\alpha}$). It was shown that for $\alpha\to 0$ this yields ideal boson-sampling, for small but non-zero amplitude it is provably computationally hard by treating the cat state as an error model, and for general $\alpha$ strong evidence was provided that the problem is hard by relating the amplitudes to a permanent-like function. Intuitively, one might expect that there is a plethora of non-Fock quantum states of light that yield computationally hard sampling problems, a question that warrants future research.

\section{How to build a boson-sampling device}

In this section we explain the basic components required to build a boson-sampling device. This device consists of three basic components: (1) single-photon sources; (2) linear optics networks; and, (3) photodetectors. Each of these present their own engineering challenges and there are a range of technologies that could be employed for each of these components. However, although boson-sampling is much easier to implement than full-scale LOQC, it remains challenging to build a post-classical boson-sampling device. While challenging, a realizable post-classical boson-sampling device is foreseeable in the near future. 

\subsection{Photon sources}

The first engineering challenge is to prepare an input state of the form of Eq.~\ref{eq:input_state}. This state may be generated using various photon source technologies. For a review of many of the photon sources see Ref.~\cite{bib:SourceAndDetectorReview}. Presently, the most commonly employed photon source technology is spontaneous parametric down-conversion (SPDC).

\begin{figure}[!htb]
\includegraphics[width=0.7\columnwidth]{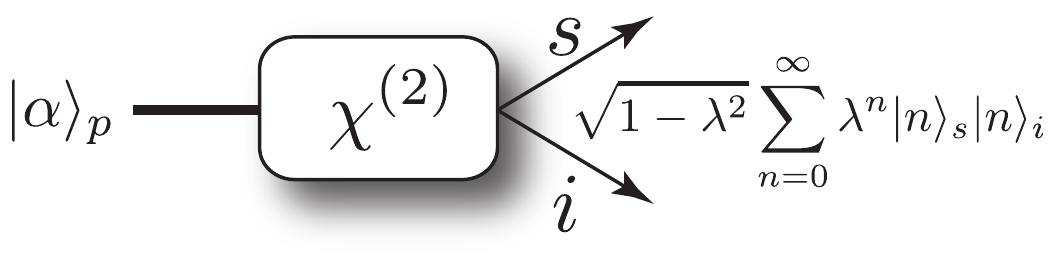}
\caption{Spontaneous parametric down-conversion (SPDC) source. A crystal with a second order non-linearity, $\chi^{(2)}$, is pumped with a classical coherent light source $\ket\alpha_p$. The source then probabilistically emits photon pairs into the signal and idler modes, including vacuum $\ket{0}\ket{0}$ and higher order terms where multiple pairs are emitted.}
\label{fig:SPDC}
\end{figure}

The SPDC source works by first pumping a non-linear crystal with a coherent state $\ket{\alpha}$ as shown in Fig. \ref{fig:SPDC}. A coherent state is well approximated by a laser source. With some probability one of the laser photons interacts with the crystal and emits an entangled superposition of photons across two output modes, the \emph{signal} and \emph{idler}. The output of an SPDC source is of the form \cite{bib:GerryKnight05},
\begin{equation} \label{SPDC}
\ket{\Psi_\mathrm{SPDC}} = \sqrt{1-\chi^2}\sum_{n=0}^{\infty}\chi^n\ket{n}_s\ket{n}_i,
\end{equation}
where $0\leq\chi\leq1$ is the squeezing parameter, $n$ is the number of photons, $s$ represents the signal mode, and $i$ represents the idler mode. For boson-sampling, we are interested in the $\ket{1}_s\ket{1}_i$ term of this superposition since we require single photons at the input of the first $n$ modes. The signal photons are measured by a photodetector and because of the correlation in photon-number, we know that a photon is also present in the idler mode. The idler photons are then routed into one of the input ports of the boson-sampling device using a multiplexer \cite{bib:migdall2002tailoring, bib:LPOR201400027, bib:ma2011experimental}.

There are several problems associated with SPDC sources, which limit the scalability of boson-sampling. The major problem is higher order photon-number terms. In the boson-sampling model we only want the $\ket{1}_s\ket{1}_i$ term, which is far from deterministic. The SPDC source is going to emit the zero-photon term with highest probability and emit higher order terms with exponentially decreasing probability. If the heralding photodetector does not have unit efficiency, then the heralded mode may contain higher order photon-number terms.

\begin{figure}[!htb]
\includegraphics[width=0.7\columnwidth]{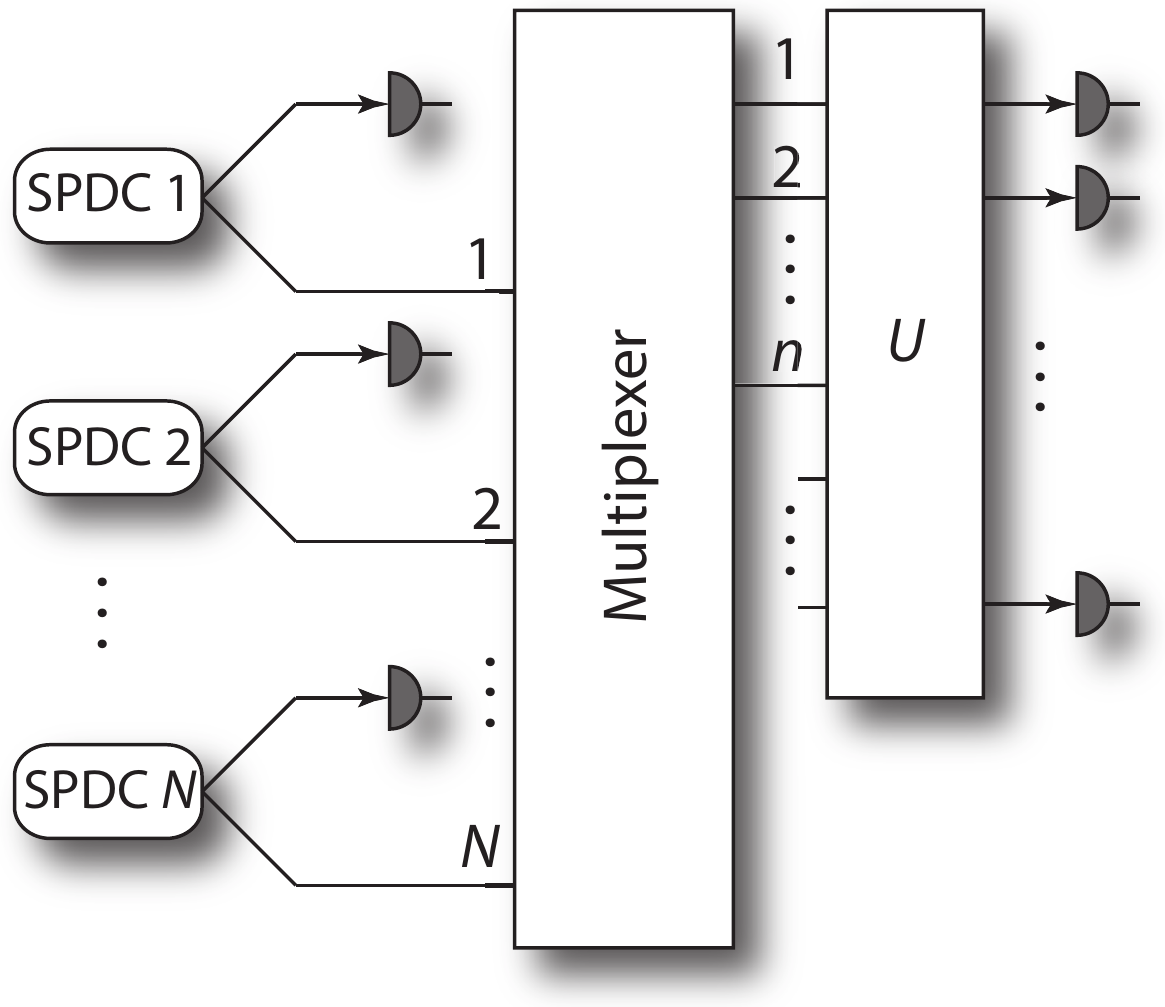}
\caption{Boson-sampling architecture using SPDC sources with an active multiplexer. $N$ sources operate in parallel, each heralded by an inefficient single-photon number-resolving detector. It is assumed that \mbox{$N\gg n$}, which guarantees that at least $n$ photons will be heralded. The multiplexer dynamically routes the successfully heralded modes to the first $n$ modes of the unitary network $\hat{U}$. Finally, photodetection is performed and the output is post-selected on the detection on all $n$ photons.}
\label{fig:multiplexing}
\end{figure}

It was recently shown by Motes \emph{et al.} \cite{bib:motes2013spontaneous} that SPDC sources are scalable in the asymptotic limit for boson-sampling using a multiplexing device. Specifically, if the photodetection efficiency is sufficient to guarantee post-selection at the output of the boson-sampling device with high probability, then the heralded SPDC photons also have asymptotically high fidelity. The boson-sampling architecture with multiplexing is shown in Fig.~\ref{fig:multiplexing}. Furthermore, it was shown by Lund \emph{et al.} \cite{bib:LundGauss} that one can do away with the multiplexer altogether, simply routing the SPDC outputs directly to the interferometer (which has become known as `scattershot' boson-sampling), which still yields an equivalently hard sampling problem, a massive experimental simplification.

Another problem is that photons from SPDC sources have uncertainty in their temporal distribution. If a boson-sampling device is built using multiple SPDC sources it is difficult to temporally align each of the $n$ photons entering the device. This is called temporal mismatch. The error term associated with this scales exponentially with $n$, yielding an error model consistent with Eq.~\ref{eq:error_model}, which undermines operation in the asymptotic limit. 

\subsection{Linear optics networks}

After the input state has been prepared it is evolved via a linear optics network, $\hat{U}$. $\hat{U}$ transforms the input state as per Eq.~\ref{eq:Utransform} and may be completely characterized before the experiment using coherent state inputs \cite{bib:PhysRevLett.73.58}. $\hat{U}$ is composed of an array of discrete elements, namely, beamsplitters and phase-shifters. A beamsplitter with phase-shifters may be represented as a two-mode unitary of the form \cite{bib:GerryKnight05},
\begin{equation} \label{eq:BS}
U_{\mathrm{BS}}(t) = \left( \begin{array}{cc}
e^{i(\alpha-\frac{\beta}{2}-\frac{\gamma}{2})}\mathrm{cos}\left(\frac{\delta}{2}\right) & -e^{i(\alpha-\frac{\beta}{2}+\frac{\gamma}{2})}\mathrm{sin}\left(\frac{\delta}{2}\right)  \\
e^{i(\alpha+\frac{\beta}{2}-\frac{\gamma}{2})}\mathrm{sin}\left(\frac{\delta}{2}\right) & e^{i(\alpha+\frac{\beta}{2}+\frac{\gamma}{2})}\mathrm{cos}\left(\frac{\delta}{2}\right)
\end{array} \right), 
\end{equation}
where \mbox{$0\leq\alpha\leq2\pi$} and \mbox{$0\leq\{\beta,\gamma,\delta\}\leq\pi$} are arbitrary phases.

For a $\hat{U}$ that implements a classically hard problem one would need hundreds of discrete optical elements. Constructing an arbitrary $\hat{U}$ using the traditional linear optics approach of setting and aligning each optical element would be extremely cumbersome. Thus, using discrete optical elements is not a very promising route towards scalable boson-sampling.

One method to simplify the construction of the linear optics network is to use integrated waveguides. Quantum interference was first demonstrated with this technology by Peruzzo \emph{et al.} \cite{bib:peruzzo2011multimode}. This technology requires more frugal space requirements, is more optically stable, and far easier to manufacture, allowing the entire linear optics network to be integrated onto a small chip \cite{bib:Politi02052008, bib:matthews2009, bib:Politi04092009}. The main issue with integrated waveguides is achieving sufficiently low loss rates inside of the waveguide and in the coupling of the waveguide to the photon sources and photodetectors. Presently, the loss rates in these devices are extremely high and thus post-selection upon $n$ photons at the output occurs with very low probability. It is foreseeable that photon sources and photodetectors will eventually be integrated into the waveguide which would eliminate coupling loss rates, substantially improving scalability.   

\begin{figure}[!htb]
\includegraphics[width=0.7\columnwidth]{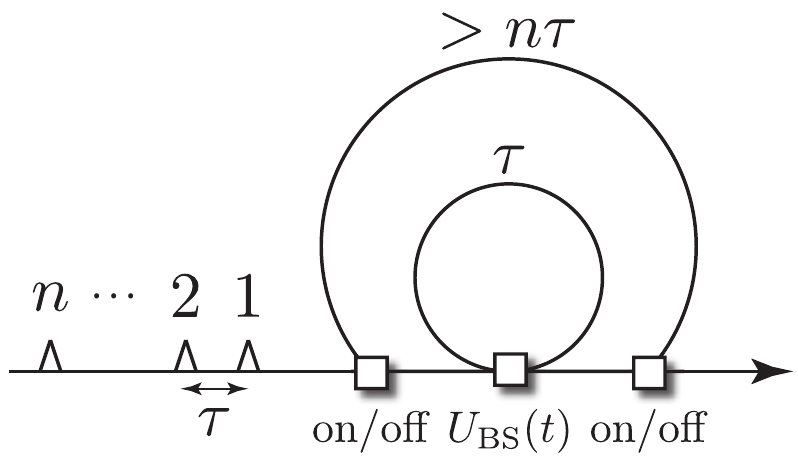}
\caption{Time-bin encoding architecture for implementing a boson-sampling device. Single photons arrive in a train of time-bins instead of in spatial modes. Each time-bin corresponds to spatial modes in the boson-sampling scheme and are separated by time $\tau$. The photon train is coupled into the loop by the first switch. The photons then traverse the inner loop such that each time-bin may interact. The first (last) photon is coupled completely in (out). The outer loop allows an arbitrary number of the smaller loops to be applied consecutively which is determined by the third switch. Finally, the photon train is measured at the output using time-resolved detection.}
\label{fig:fiber_loop}
\end{figure}

Another potential route to simplifying the linear optics network is to use time-bin encoding in a loop based architecture \cite{bib:motes2014scalable}. The major advantage of this architecture is that it only requires two delay loops, two on/off switches, and one controllable phase-shifter as shown in Fig.~\ref{fig:fiber_loop}. This possibility eliminates the problem of aligning hundreds of optical elements and has fixed experimental complexity, irrespective of the size of the boson-sampling device. A major problem with this architecture however is that it remains difficult to control a dynamic phase-shifter with high fidelity at a rate that is on the order of the time-bin width $\tau$.

\subsection{Photodetection}

The final requirement in the boson-sampling device is sampling the output distribution as shown in Eq. \ref{eq:outputState}. With linear optics this is done using photodetectors. For a review on various types of photodetection see Ref.~\cite{bib:SourceAndDetectorReview}.  

There are two general types of photodetectors \textemdash photon-number resolving detectors and bucket detectors. The former counts the number of incident photons. These are much more difficult to make and more expensive in general than bucket detectors. Bucket detectors, on the other hand, simply trigger if any non-zero number of photons are incident on the detector. As discussed earlier, in the limit of large boson-sampling devices, we are statistically guaranteed that we never measure more than one photon per mode, since the number of modes scales as \mbox{$m=O(n^2)$}. Thus, bucket detectors are sufficient for large boson-sampling devices, a significant experimental simplification compared to universal LOQC protocols. 

Many photodetector designs use superconductivity to measure photons. Superconductivity is an extreme state where electrical current flows with zero resistance. It occurs in conductive materials when a certain critical temperature is reached. This critical temperature is far from occurring naturally on Earth and thus high-tech and expensive lab equipment is required. For many materials this temperature is close to absolute zero. Such extreme conditions are required for detecting single photons because single photons are themselves extreme. They are after all the smallest unit of light.

\begin{figure}[!htb]
\includegraphics[width=0.45\columnwidth]{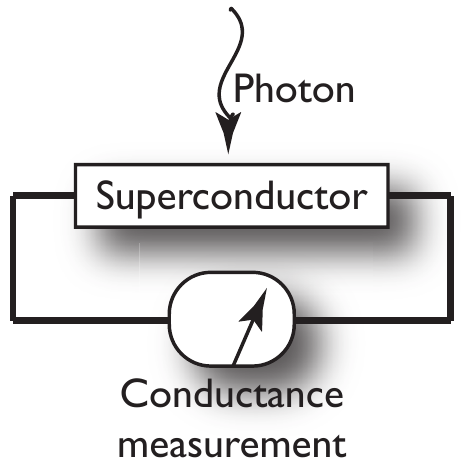}
\caption{Basic design for a superconducting photon detector. A current is passed through a superconductor and the conductance is monitored. Photons impart energy on the superconductor that is just cooled to its critical temperature. This added photon energy causes a measurable change in the conductance allowing for the detection of photons.}
\label{fig:superconducting_detector}
\end{figure}

While there are several variations of superconductive photodetectors, they tend to work similarly to the one shown in Fig. \ref{fig:superconducting_detector}. The idea is that a superconductor is cooled to a point just below its critical temperature. Current is then applied through the superconductor which experiences zero resistance. If there is no resistance, then there is no voltage drop across the superconductor and our conductance measurement reads infinity. Then the photon or photons that are to be measured will hit the superconductor and be absorbed. Each photon that is absorbed by the superconductor imparts energy $h\nu$ onto it, where $h$ is Plank's constant and $\nu$ is the frequency of the photon. This heats the superconductor above its critical temperature. The conductance measurement will then change according to the absorbed photon, thus informing the measurer that a photon was detected. This scheme may be able to count several photons since the conductance will change proportionally to the number of absorbed photons. However, if too many photons are absorbed all properties of superconductivity are lost and thus number-resolution is lost.

Photodetectors may be used to help overcome the problem of temporal mismatch. Such detectors must have the ability to record the time at which the photon arrived. If we post-select upon detecting all $n$ output photons in the same time-window $\Delta t$ then we can assume that their temporal distribution overlaps sufficiently to yield a classically hard sampling problem. This method however is not reliable for scalable boson-sampling. If the temporal distributions are not sufficiently overlapping, then the probability of post-selecting all $n$ photons in the same time-window decreases exponentially with $n$. However, if the sources are producing nearly identical photons in the time domain then this method would be a practical cross check.

As the distinguishability of photons varies the complexity of sampling the output distribution also varies. A theoretical framework was developed by Tillmann \emph{et al.} \cite{bib:tillmann2014BS} that describes the transition probabilities of photons with arbitrary distinguishability through the linear optical network. The output distribution of boson-sampling with distinguishable photons is then given by matrix immanants, thus affecting the computational complexity of the output distribution. They also test this experimentally by tuning the temporal mismatch of their input photons. This boson-sampling experiment is unique in that it is the first to use distinguishable photons at the input. 

\section{Conclusion}

In this chapter we have given an introduction to the rapidly evolving field of passive linear optical quantum computation, a new model of quantum computation that, while not universal, nevertheless can carry out efficiently \textemdash at least! \textemdash the interesting mathematical problem of boson-sampling. Since a boson-sampling output is strongly believed to be inefficient to verify on a universal quantum computer, much less a classical computer, the passive linear optical approach to quantum computation really is something new and quite different than the usual quantum computer idea. While at the present the boson-sampling problem does not have any known practical uses, it nevertheless provides us a new window into the hidden computational power of quantum mechanical devices. 

Often researchers in the field of quantum computation are asked by colleagues in other fields, or the popular press, ``When will the quantum computer be built?'' The answer depends greatly upon what exactly it is you mean by a `quantum computer'. We turn this question around and ask our colleagues in the field of computer science, ``Well first you tell us when the classical computer was built.'' One would think that at least this would have a universally agreed upon answer, but that is not the case. Many will reply that it was the ENIAC, a digital machine that came online in 1945 that was Turing complete, reprogrammable, electronic, digital, and had a memory. Originally designed to compute US Army artillery tables it was quickly turned to simulating H-bomb explosions. Others hold out for the Atanasoff\textendash Berry Computer (ABC) that came online in 1941 and which was also electronic, digital, and programmable, but not universal and lacked the equivalent of today's RAM. To muddy the waters, some implore that we consider the British electronic Colossus computers that were first constructed in 1943 but only declassified in the 1970s. However they were not at time of operation Turing complete but could apparently have been made so. The list goes on and on \cite{ceruzzi}. Why no agreement on when the classical computer was built? It is because experts in the field disagree on exactly what is a classical computer. Some even hold out for the all-mechanical Babbage machine of the 1800s. All we can say is that in 1850 the classical computer certainly had not been built and by 1950 it certainly had but that nobody can agree on a precise date or a precise computer. 

We should not expect any more or less of the future history of the quantum computer. We have ion trappers painstakingly assembling a universal quantum computer a qubit at a time. But then we have the company D-Wave the makes non-universal quantum machine that nevertheless implements a watered-down version of adiabatic quantum computing, called quantum simulated annealing, that appears to show some polynomial improvement on particular problems such as structured search or pattern matching algorithms. Then here comes the passive linear optical device that implements boson-sampling efficiently. What to think about that? The device is not universal, in that it cannot solve every problem a universal quantum computer can solve, but nevertheless it can solve the boson-sampling problem efficiently. In some sense the passive linear optics model of computation is as powerful as a universal quantum computer, but only when it comes to this one problem. It is a bit like the ABC machine, which was not universal but really good at solving linear sets of equations, or like the Colossus machines that were also not universal but really good at cracking a particular type of German cipher called Enigma. Those problems were certainly useful and it remains to be seen if our non-universal passive linear optical machine has the ability to solve some useful problems efficiently as well \cite{dowlingbook}. But in the meantime it is certainly interesting to be working at this interface of quantum optics and quantum computation where the work of AA and others have provided us with a totally new ball game. Who knows what other secrets lay hidden in simple interferometers? All we can say for sure here is that there is a great future in photons. 

To recap this chapter, in Section I, we have reviewed universal quantum computation in general and then particularly with photons, the so-called linear optical approach. While universal it is unwieldy given the overhead associated with active optical elements. Far simpler is the passive linear optics interferometer with no moving parts, and to such a machine there fits quite nicely the boson-sampling problem, as outlined in Section II. Here we also discussed the error tolerance required for such a machine in order for it to show a computational speedup. In Section III we take a detour through quantum computer complexity theory and discuss just what experiments with boson-sampling do and do not prove in that realm. Specifically, we argue that, contrary to a popular claim made in the literature, boson-sampling cannot disprove or even provide evidence against the Extended Church-Turing thesis. The original boson-sampling paradigm of AA admitted only single photons as input. In Section IV we discuss what happens if that assumption is relaxed. We surmise that any passive linear optical network (that implements a random unitary from the Haar class) fed with negative Wigner function states yields a hard sampling problem. In Section V we review how the quantum mechanics out there actually build such devices and we delve into a number of hardware and implementation issues. Finally in this Section VI the section refers to itself in true G\"{o}delian fashion by concluding. Now if we were to take the set of all book chapter conclusions that conclude themselves\ldots

%
%

\begin{acknowledgments}
KRM and PPR would like to acknowledge support from the the Australian Research Council Centre of Excellence for Engineered Quantum Systems (Project number CE110001013). JPD would like to acknowledge support from the Air Force Office of Scientific Research and the Army Research Office. BTG would like to acknowledge support from the National Physical Science Consortium \& National Institute of Standards and Technology graduate fellowship program. JPO would like to acknowledge support from a Louisiana State University System Board of Regents Fellowship. 
\end{acknowledgments}

%
%

\bibliography{bibliography_6-23}

\end{document}